\documentclass[fleqn,10pt]{wlscirep}
\usepackage[utf8]{inputenc}
\usepackage[T1]{fontenc}
\title{Bias Impact Analysis of AI in Consumer Mobile Health Technologies: Legal, Technical and Policy}

\author[1,*]{Kristine Gloria}
\author[2]{Nidhi Rastogi}
\author[3]{Stevie DeGroff}
\affil[1]{The Aspen Institute, Aspen Digital, Washington DC, USA}
\affil[2]{Rochester Institute of Technology, Rochester, USA}
\affil[3]{Independent Researcher, USA}

\affil[*]{Kristine.Gloria@aspeninstitute.org}

\begin{abstract}
Today's large-scale algorithmic and automated deployment of decision-making systems threatens to exclude marginalized communities. Thus, the emergent danger comes from the effectiveness and the propensity of such systems to replicate, reinforce, or amplify harmful existing discriminatory acts. Algorithmic bias exposes a deeply entrenched encoding of a range of unwanted biases that can have profound real-world effects that manifest in domains from employment, to housing, to healthcare. The last decade of research and examples on these effects further underscores the need to examine any claim of a value-neutral technology. This work examines the intersection of algorithmic bias in consumer mobile health technologies (mHealth). We include mHealth, a term used to describe mobile technology and associated sensors to provide healthcare solutions through patient journeys. We also include mental and behavioral health (mental and physiological) as part of our study. Furthermore, we explore to what extent current mechanisms - legal, technical, and or normative - help mitigate potential risks associated with unwanted bias in intelligent systems that make up the mHealth domain. We provide additional guidance on the role, and responsibilities technologists and policymakers have to ensure that such systems empower patients equitably. 
\end{abstract}
\begin{document}

\flushbottom
\maketitle
%
%


\section{Introduction}

The systematic or discriminatory exclusion of marginalized communities is not a new phenomenon. What is new, however, are today's large-scale algorithmic and automated decision-making systems that may consistently deploy and repeat such discriminatory practices generating unfair outcomes\cite{01}. The term "algorithmic bias" provides the construct and the language to examine and articulate these effects more astutely. Algorithmic bias exposes a deeply entrenched encoding of a range of unwanted biases that can have profound real-world effects that manifest in domains from employment, to housing, to healthcare. The last decade of research and examples on these effects further underscores the need to examine any claim of a value-neutral technology. Author and researcher Brian Christian write: 
\par
\textit{"Our human, social, and civic dilemmas are becoming technical. Furthermore, our technical dilemmas are becoming human, social, and civic. Our successes and failures in getting these systems to do 'what we want,' it turns out, offers us an unflinching, revelatory mirror.\cite{02}."}
\par
Thus, the emergent danger comes from such systems' effectiveness and propensity to replicate, reinforce, or amplify harmful existing discriminatory acts~\cite{03}. The following adds to a growing corpus of research into potential challenges of algorithmic systems and offers potential mitigation strategies. Specifically, this work examines the intersection of algorithmic bias in consumer mobile health technologies (mHealth). Deloitte defines mHealth as 
\par
\textit{
"Any use of mobile technology (e.g., networks, sensors, mobile devices) to provide healthcare solutions throughout the patient journey~\cite{04}."}
\par
Within mHealth, we also pay close attention to the growing sub-category of digital mental health and behavioral healthcare interventions. Walsh et al. define "behavioral healthcare" to include emotional, mental, and social factors as well as behaviors to prevent illness (e.g., avoiding substance abuse) and promote wellness (e.g., exercise)\cite{05}. We explore to what extent current mechanisms - legal, technical, and/or normative - help mitigate potential risks associated with unwanted bias in intelligent systems that make up the mHealth domain. We provide additional guidance on the role and responsibilities technologists and policymakers have to ensure that such systems empower patients equitably.

\section{What is Algorithmic Bias?}
There are several interpretations for algorithmic Bias. For our purposes, we unpack the term into two main components: Algorithm and Bias. Data \& Society defines an algorithm as: 
\par
\textit{"A set of instructions for how a computer should accomplish a particular task. Combining calculation, processing, and reasoning, algorithms can be exceptionally complex, encoding for thousands of variables across millions of data points\cite{06}}." 
\par
In the current scenario, algorithms increasingly learn from massive data. This data may be provided by humans or collected from machines, or a combination of the two. Bias can be defined as "outcomes which are systematically less favorable to individuals within a particular group and where there is no relevant difference between groups that justifies such harms.\cite{07}''

Algorithmic Bias is also a repeatable error that creates unfair outcomes for specific groups and often manifests within large-scale technology systems. While human biases can influence algorithms that result in discriminatory outcomes, not all algorithmic outcomes are harmful. As companies and government institutions deploy an automated decision-making system to gain efficiencies and scale, this challenge comes to bear. Companies like IBM have created comprehensive frameworks and tools to help identify and mitigate AI systems to prevent discriminatory outcomes. Most recently, the IBM Policy Lab published a risk-based AI governance policy framework based on accountability, transparency, fairness, and security to apply more narrowly-tailored policy approaches for addressing discrete harms within specific spaces (e.g., criminal justice and health care)~\cite{08}.

\subsection*{When, where, and how is the bias introduced?}
Research suggests several entry points by which unwanted bias may enter into the development cycle of an algorithm and the resulting technical system. This includes instances such as the very design of the study itself to the most commonly referenced data development process. Data science literature identifies five main stages in data development: 
\par
\textit{define and collect; label and transform; analyze and insight; modeling; and deployment~\cite{09}}. To illustrate, data collection is a systematic approach grounded in solving the problem. And this first stage is paramount for any attempt towards responsible data usage. Data collection should be even more principled and moral when it relates to personal information~\cite{10}. 
\par
\begin{figure}[h]
\caption{Where Bias can be introduced}
\centering
\includegraphics[width=.8\textwidth]{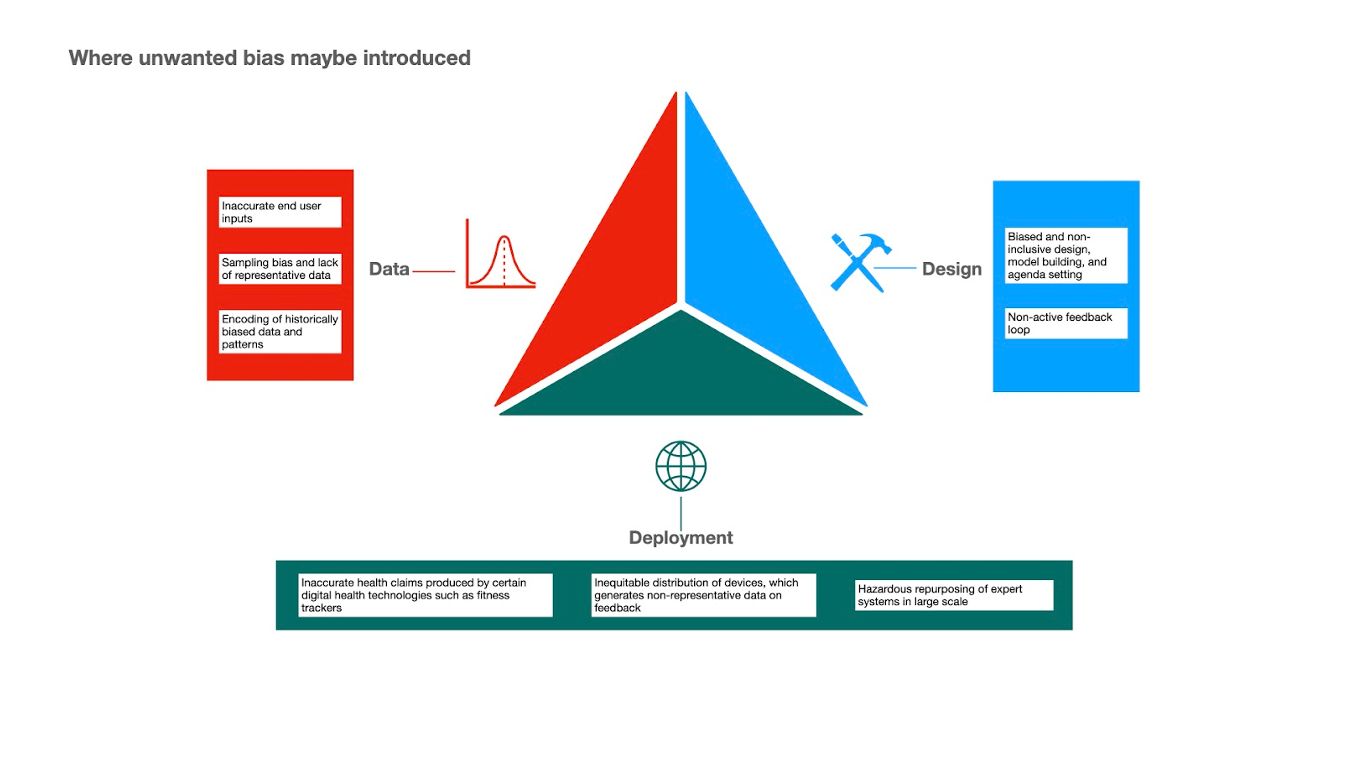}
\label{picture1}
\end{figure}

Unfortunately, bias may enter at this stage for various reasons such as a lack of representative data, data labeling, see Figure~\ref{picture1}~\cite{11}, and or cost-prohibitive access to data. For mhealth solutions, the challenge also reflects a level of stigma or under-reporting (e.g., leading to under-or over-reporting) and a lack of unreliable biomarkers~\cite{12}.

\par
Unwanted bias on the data level, however, can be addressed. In statistics and informatics, one valuable exercise is assessing data quality for completeness and comprehensiveness, accuracy and precision, timeliness, and relevance. However, the problem is that statistical correctness and/or parity is only part of the equation~\cite{13}. Another layer of complexity arises as part of the datasets' labeling, classification, and transformation process. In the last several decades, information architecture has evolved from a more stringent set of hierarchical schema to ones that now allow for more flexibility, such as relational and object-oriented databases and knowledge graphs. However, inherent in any data organization remains the practice of categorization or classification. Any classification system embodies a dynamic compromise that encapsulates what belongs and does not~\cite{14}. Mona Sloane writes, 
\par
\textit{
"The notion that truths, values, \& culture can be codified, quantified, \& extracted from how individuals behave under artificial conditions assumes that these amorphous concepts are external to lived experiences or social context~\cite{15}."}
\par
As we have argued previously, algorithmic bias-- at its core-- is about power. It systematically codifies what knowledge, models, and truth are most advantageous. It raises questions about where power is vested and privilege to designate knowledge components and narratives and how deeply these human judgments are entrenched in myriad technical systems. The lens of algorithmic bias enables researchers to critically discern between intentional and unintentional biases, which we will define further in the paper. Moreover, while research clarifies that bias may enter at various points of the development of a system, it is rarely the result of one discrete action or decision. Instead, the concept illuminates the role a set of actors play in shaping the algorithm and the system they operate. Solon and Selbst write:
\par
\textit{
"While useful, data is not the solution to all problems. Data is used predictively to assist decision-making; it can affect the fortunes of whole classes of people inconsistently in unfavorable ways. Sorting and selecting the best or the most profitable candidates means building a model with the power to select winners and losers. Caution is required, lest we see this process resulting in adverse outcomes disproportionately concentrated for historically disadvantaged groups, which is akin to discrimination~\cite{17}."}
\par
While the "data miners" class is often used to describe the class of humans responsible for designing and implementing data collection processes, they may not necessarily be carried out by the same parties. The data collection and analysis task usually follow defining a hypothesis - what do we want to learn or what are we interested in predicting by using the data? Thus, in some cases, the narrative of the task is already defined even before the data collection process begins. Therefore, it is critical to recognize that bias can be introduced early in the process, intentional or not.

\section{What are consumer mobile health technologies?}
It can be argued that the medical and healthcare domain offers an exemplary environment - with its experimental and data-centric nature - for adopting and deploying large-scale algorithmic systems. Applied machine learning, which draws upon a constellation of statistics, algorithms, and data models, can be found in applications such as clinical diagnostics to insurance allocation of health care for patients. There is, however, less focus on the implication of unwanted bias in expert systems related to consumer health technologies, which include multiple solutions from wearables to virtual care. To some, these solutions enable and empower consumers to take charge of their health outcomes proactively. To others, consumer health technologies perpetuate concerns that surfaced through the critical lenses of big data and surveillance capitalism. Moreover, recent work questions the accuracy and efficacy of health claims produced by certain digital health technologies such as fitness trackers~\cite{18}. We articulate a set of potential harms below. 
\par
Despite such critiques, the adoption of commercial, consumer-driven health technologies is on a positive growth curve. In 2019, the estimated global digital health market was approximately \$175 billion (USD). By 2025, the estimated market size will approach nearly \$600 billion (USD)~\cite{19}. Some early indicators suggest that the COVID-19 global pandemic accelerated consumers' need and adoption of digital health tools. As Morris Panner notes in Forbes~\cite{20}, "The pandemic confronted consumers with taking care of themselves proactively, and remote and virtual services showed the range of care options because of technology ."For digital behavioral technologies, the U.S. market is projected to grow from \$77.62 billion (USD) in 2021 to \$99.40 billion (USD) in 2028~\cite{21}. This upward trend in the adoption of mobile digital health tools tracks with an (uneasy) set of statistics that indicate that approximately a billion people are enduring some kind of mental disorder, three million people succumb to alcoholism every year, and one person dies every 40 seconds by suicide~\cite{22}. The key takeaway is that digital health technologies, telehealth, wellness trackers, and (or) mental health apps are here to stay. The COVID pandemic, at the minimum, helped sharpen this view. For our purposes, we explore whether the mHealth domain has been adequately assessed for the impact of unwanted bias and to what extent our current guardrails (e.g., regulation) address disparate or discriminatory outcomes. 

\subsection*{Why we need further study}
In some regards, the medical and health industry has set the standard for various practices that prioritize protecting patients' personal information. In the U.S., the Health Insurance Portability and Accountability Act (1996) ("HIPAA") regulates personal health information collected and used by covered entities and business associates. Its intent and success range from preventing fraud to streamlining administrative healthcare functions. Underlying it all is a core assumption in digital privacy that health data be treated and classified as especially sensitive. Therefore, medical and health professionals that collect and use a patient has covered digital health information must often meet a higher standard than general digital solutions. However, HIPAA's scope is limited. The consumer mobile health space is not subject to HIPAA and does not share the same burden or level of scrutiny. Instead, commercial mHealth solutions emulate the same rules and practices of technologies such as social media networks, which are largely self-regulated. Specifically, HIPAA applies only to data transferred to or from "covered entities" (e.g., health care clearinghouses and providers). mHealth apps downloaded and used by consumers, without the direct association of a covered entity, do not fall under HIPAA~\cite{23}. Thus, the law seeks to ensure the confidentiality of patient data. It does not directly address fairness or inclusivity of analyses or recommendations that rely on that data.
\par
The expansive growth of the mHealth market underscores its potential for more frequent, granular, and widely accessible health interventions across various needs. While mHealth solutions do not operate as a single large-scale technology, the combination of replicated AI-driven recommendations and the sheer volume of adoption creates a large-scale network of technologies that may amplify discriminatory outcomes. Specifically, the decisions that inform the design and build of any mHealth intervention can impact people's health and well-being. In addition, as these apps leverage machine learning models to create personalized or adaptable health interventions, biased data or biased assumptions (intentional or not) can exacerbate the risk of bias or make it more challenging to discover and mitigate. Most recently, Swiss Re's annual Systematic Observation of Notions Associated with Risk (SONAR) warned insurers. It cautioned that users could face losses from litigation against manufacturers in cases where inappropriate advice provided by health technology and apps, such as wearable devices, resulted in user injury or other patient impairments~\cite{24}. The following section offers a specific use case to help illuminate additional challenges and implications of algorithmic bias in mHealth solutions.

\subsubsection*{The potential of bias in mHealth}
As with any technology system, mHealth apps reflect a set of decisions - from defining the app's objective to user-centric design to validating models - that affect how the app operates, who uses it, its outputs, and its potential effect on a user's health. At these multiple decision points, bias may enter - intentionally or not. In some instances, mHealth solutions are developed by teams that do not represent or have the lived experience of the end-user, which may result in implicit bias. In another example, bias may enter when using incomplete or under-represented data in the models. This is further complicated as more apps increase the use of machine learning, making it challenging to discover and correct bias. The following section expands on potential insertion points of bias.

\section{Use Case - COVID-19 Trackers}
Marginalized communities have experienced the most damaging effects of the COVID-19 pandemic~\cite{25}. Machine learning (ML) models, which identify patterns from large datasets, are an intrinsic part of the health informatics toolkit to fight COVID-19 disease. While algorithms are not inherently biased, training them with biased distribution towards a specific section of society results in biased predictions made by the model. Further, indiscriminately deploying ML thus risks amplifying the adverse effects of the pandemic on vulnerable groups, intensifying health inequity. This section discusses the bias and discrimination in the design and deployment of mobile apps that analyze their data using ML models and are prone to influencing biased healthcare initiatives and publishing prejudiced study conclusions.
\par
\subsection*{Utilizing mobile apps to inform COVID-19 response}
Numerous mobile phone applications (called apps) are available on the Apple store and Google Play for monitoring the effectiveness of COVID-19 interventions. The collected data from these apps have been widely used and continue to assess potential spatial and temporal spread drivers, including the support for local administrative contact tracing efforts. Unmistakably, mobile phone data remains one of the best data collection methods for studying large-scale population behaviors~\cite{26}. This data has heavy utility in analyzing population patterns such as changes in mobility, clustering of social contacts, and symptom tracing in both high and low-income settings to inform the response to Covid-19. Without a doubt, the mobile app mined data is an essential part of the COVID-19 response.
\par
Nonetheless, a careful understanding of the behaviors and populations they capture should be considered alongside the data and analysis. Specifically, before guiding and evaluating the response from mobile data, it is necessary to capture the sources of potential bias and implications of bias. Through various social programs, mobile phones have been provided to disadvantaged and vulnerable communities to support monitoring the COVID-19 spread. We should consider identifying the presence of any strong bias towards these communities~\cite{27}.

\subsection*{Concerns around the usage of mobile apps}
\begin{figure}[ht]
\centering
\includegraphics[width=0.6\linewidth]{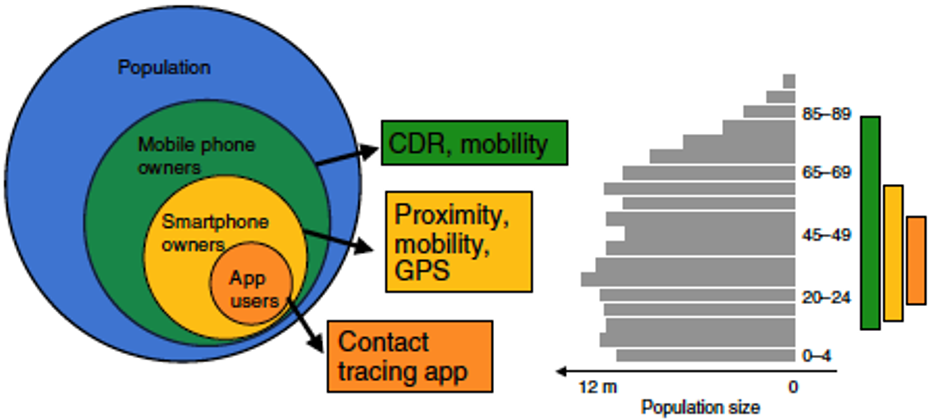}
\caption{``The use of mobile phone data to inform analysis of COVID-19 pandemic epidemiology~\cite{28}.''}
\label{fig:mobileData}
\end{figure}

This section covers privacy and bias concerns regarding mobile phone data usage. Mobile phone operators use unique identifiers for all their service subscribers. Call Detail Records (CDRs), messages collected by these mobile phone operators, contain identifying information about the subscribers, such as the timestamp of the call or message, GPS location, and the unique subscriber ID~\cite{28}, see figure~\ref{fig:mobileData}. Mobility patterns and metrics of the subscribers can be easily estimated using this data for characterizing large, population-level mobility patterns. The frequency of trips between two locations in a given period reveals temporal trends during regular and holiday periods. The import risk can be understood using this information from areas with outbreaks to new, vulnerable but without sustained transmission. Retracing a likely introduction and spread and using the occurrence of such information can inform future projections of disease risk and implement travel restrictions.

\subsection*{Data Collection Bias}
There have been attempts to identify risks of infection as an outcome of Covid-19 in several investigations based on observations. Confounding bias can arise when researchers restrict the analysis of a confounding variable~\cite{29}. In COVID-19 studies, confounding bias may emanate from restricting analyses to those hospitalized, tested for active infection, or those who volunteered their participation in a large-scale study. The relationships between any variables related to hospitalization will be distorted among hospitalized patients compared to the general population. Confounding bias-induced associations are properties of the representative data, not an individual patient. Therefore, such associations cannot be relied upon to determine individual-level causal effects. Associations based on gathered COVID-19 datasets may not reflect patterns in the population of interest if researchers look only within the subset of hospitalized patients. Likewise, due to a shortage of internal validity, such effects may not be accurate within the dataset.
\subsection*{Selection Bias}
Selection bias can occur when effect modifiers are distributed differently in the sample than in the population, thus causing effects to differ between the two~\cite{30}. Datasets often under-sample those with irregular or limited access to the healthcare system, such as socioeconomically disadvantaged groups. Limited access to technology (smart sensors) and digital access is not the only contributor to creating the potential for biased datasets. The continued association of electronic health records with pervasive sensing data will only exacerbate underrepresentation. Therefore, when biases from existing practices affect those datasets, the algorithmic models they generate will reproduce inequities.
\subsection*{Model Bias}
AI systems do not suffer only from under-representation and patterns of discrimination as sources of bias. Long-held institutional racism and the implicit - often unconscious - biases of AI developers and users might influence AI design and deployment choices. As a result, discrimination and prejudice have been integrated into building AI models, as the designers rarely represent the marginalized group\cite{31}. Therefore, there is a misrepresentation of problem formulation, the performance of data-centric analysis, and the kind of technologies to be developed for the group that most needs it. COVID-19 exacerbated this concern, as the urgency to find solutions and the institutional hierarchies in decision-making are at cross purposes with consensus-building mechanisms. The diligence needed to ensure oversight and community involvement in setting the agenda may be missing too. For modeling ML models, during the stage of defining target variables and their quantifiable proxies, any hidden and deep biases perpetuating from the designers, developers, and researchers might introduce structural health inequalities and injustices in the model via label determinations (that is, choices made in the specification of target variables) that cannot capture complexities of the social contexts of discrimination\cite{32,33}. Implicit biases in the source data can have a ripple effect when data uploaded to a public repository are further used for research and study. For example, more interesting, unusual, or severe cases of COVID-19 likely appear in publications.
\subsection*{Implicit Bias}
There is also discussion of an implicit bias occurring when problems are framed from the perspective of majority groups, using implicitly biased assumptions about data, and using outputs of AI tools to perpetuate biases, either inadvertently or explicitly. Data scientists acknowledge that diverse data alone would not guarantee the elimination of data bias. Ultimately, people are responsible for data collection and use, raising concerns about the lack of diversity among developers and funders of AI tools, causing implicit bias. More diverse data, disaggregated by societal hierarchies and clusters, is needed to determine the extent to which underrepresented groups with similar complaints and disease severity as majority groups have received similar healthcare.
\newline
\textbf{\textit{What is the harm?}} What do we mean when discussing harm resulting from unwanted bias in mHealth products? Unwanted algorithmic bias is not always intentional or harmful. Instead, as discussed above, unwanted bias is a compilation of decision points within a system's design and build. It can be introduced at various decision points, from system design to data collection to algorithmic inferencing. However, if unchecked, large-scale harm can indeed manifest. While not meant to be exhaustive, we posit three basic types of harms that can impact both the individual and collective levels: 1) data privacy, 2) inaccuracy harms, and 2) inequity and access. 
\par
The first potential harm is on the data privacy side. As with any digital technology, data privacy harms can arise with the volume of data collected, storage (e.g., length of time, security), and its use or misuse (or contextual integrity) by the company or third parties. Second, mHealth apps that market specific interventions to address symptoms or behaviors, such as anxiety or hypertension, may not reflect the latest scientific evidence or recommended interventions. Moreover, mHealth solutions grounded in medical science may risk inheriting biases (e.g., data sampling and selection bias) implicit in the scientific study. Lastly, while the mHealth market allows for adoption at scale by often circumventing the obstacles of traditional medical payment systems (e.g., health insurance), such apps in low-income populations face unique barriers, including a lack of health literacy. This inequitable access to and use of mHealth apps is problematic, especially if a system adjusts its models or algorithm based on user population data inputs.

\begin{table*}[]
\centering
\caption{Type of harm and the corresponding exemplar impact on an individual and on the society.}
\label{tab:MITREattackCerberus}
\resizebox{0.9\textwidth}{!}{%
\begin{tabular}{|p{3cm}|p{8cm}|p{8cm}|}\hline
\textbf{Type of Harm} & \textbf{Individual}& \textbf{Societal}\\ \hline
Data Privacy &	E.g., breach of personally identifiable information; misuse of information &	E.g., implications for privacy laws and norms\\ \hline
Inaccuracy &	E.g., inaccurate diagnostics or behavioral intervention; loss of agency in determining care & E.g. normative behavioral setting through mass adoption of an app and its recommended behavioral interventions\\ \hline
Inequity and Access & E.g., lack of health literacy, digital literacy & E.g., propagating biased data sampling and inference \\ \hline
\end{tabular}%
}
\end{table*}

\section{Mitigation Strategies and Current Guardrails}
In this section, we bring forth promising approaches for addressing bias in algorithms. We emphasize the necessity of a multi-prong approach which includes legislation, audits, ethical frameworks, and grassroots activism and lobbying. This is not a comprehensive list. Instead, the following highlights work garnered significant attention over the past few years. 

\subsection*{Technology strategies}
From a technology availability standpoint, we must incorporate diligent, deliberate, and end-to-end bias detection and mitigation protocols. Multiple toolkits, such IBM's AI Fairness 360\cite{34} (AIF360-open source Python toolkit for algorithmic fairness) or Amazon's SageMaker Clarify\cite{35}, and or Microsoft's InterpretML\cite{36} are available to help facilitate evaluations of algorithms, identify bias in datasets, and explain predictions. We also see growing interest in additional efforts on explainable AI, AI system factsheets,\cite{37} datasheets for datasets\cite{38}, bias impact statements\cite{39}, and many others. Recently, Google and the Partnership for AI created and published a framework for auditing AI systems targeted at engineering teams, named Scoping, Mapping, Artifact Collection, Testing, and Reflection (SMACTR). Unlike previous audit processes, this work provides an end-to-end framework to be applied throughout the internal organization development life-cycle. The SMACTR audit is informed by other fields "where safety is critical to protect human life, such as aerospace and health care, which now carries out audits as part of the design process\cite{40}." Additionally, and equally critical, is the inclusive community development, interdisciplinary knowledge, and ethics embedded in AI project teams to help identify and remediate potential factors that could be discriminatory.

Similarly, data-gathering practices must integrate a quantifiable understanding of the social factors of disparate vulnerability to COVID-19. This will allow the integration of data on socioeconomic status with other ethnicities and sensitive data. This will allow scrutiny of subgroup differences in processing results.
\subsection*{Normative Strategies}
Concerns of unwanted bias in large-scale automated decision-making systems have surfaced in various best practices that span across verticals and are stakeholder specific. For our purposes, we highlight work produced by the Center for Democracy and Technology, which focuses on several critical questions for developers to ask throughout the development process of health-based apps. These include:
\begin{enumerate}
    \item Are health interventions in the app evidence-based?
    \item What type of user are you designing for, and how will they access the app?
    \item Is your team diverse, and if not, how will you represent your intended users?
    \item Is your app accessible to users with low vision, blindness, hearing, cognitive, or motor impairments?
    \item Where did your model "training" data come from, and whom does it represent?
    \item How much error is acceptable in your app?\cite{41}
\end{enumerate}
These questions, a sample of the more extensive set, are constructive at the beginning of any development cycle. To address the entire product life cycle, combining these pre-development questions with a post-dev auditing mechanism would ensure a thorough examination of bias. One point of further study is the economic feasibility of implementing such a complex assessment program.
\subsection*{Policy and regulatory strategies}
Another critical prong for mitigating unwanted biases is through public policymaking and regulation. As with consumer privacy regulation, European entities have been the first movers regulating AI. The General Data Protection Regulation (GDPR) attempts to address concerns around automated decision-making by imposing explainability, transparency, and accountability requirements on AI decision systems. In April 2021, the European Commission published its proposed regulation's Artificial Intelligence Act (AIA). The Commission bases regulation and obligations on a risk scale, with high risk, requiring human oversight, risk assessments, and data management for high-risk applications that impact the health and safety of individuals. The AIA states explicitly the need for high-quality data sets for the development and testing of high-risk AI, noting that \textit{"the European health data space will facilitate non-discriminatory access to health data and the training of artificial intelligence algorithms on those datasets, in a privacy-preserving, secure, timely, transparent and trustworthy manner, and with an appropriate institutional governance\cite{42}."}
\par
In the U.S., the Federal Drug Administration (FDA) and the Federal Trade Commission (FTC) serve as potential watchdogs for emerging mHealth technologies. The FDA's mandate is limited to medical devices, something \textit{"intended for use in the diagnosis of disease or other conditions, or in the cure, mitigation, treatment, or prevention of disease . . . or . . . intended to affect the structure or any function of the body of man . . .\cite{43}."} However, Recent FDA action has focused on related software which may implicate mHealth. For example, in 2020, the FDA introduced the Digital Health Software Pre-certification (Pre-Cert) Program to \textit{"inform the development of a future regulatory model that will provide more streamlined and efficient regulatory oversight of software-based medical devices\cite{44}".} The pilot program shifts its approach from the traditional FDA medical device review by emphasizing the work of a developer responsible for building the software or the digital health technology rather than the end product. The intent is to establish a regulatory framework that is agile and responsive to consumer concerns without disrupting the continuity of service or compromising user safety.
\par
The FTC has historically regulated data privacy and security practices through its authority under Section 5 of the FTC Act, which prohibits unfair or deceptive acts or practices in consumer products. This has included the data used in consumer-facing AI applications. Specific to mHealth, a September 2021 FTC policy statement affirmed the agency's position that mobile health applications and devices which collect or use consumer health information are subject to the Health Breach Notification Rule\cite{45}. Additional FTC guidance on the use of AI has emphasized scrutinizing both data sets used by AI as well as potential discriminatory impact, transparency, accountability, and clear communication with consumers regarding the use of data in AI systems. 
\par
In addition to government regulation, ethical principles, such as the OECD AI Principles\cite{46}, the EU's Ethics Guidelines for Trustworthy AI\cite{47}, and Canada's Directive on Automated Decision-Making\cite{48},  provide helpful guidance for articulating high-level value statements which can inform the product development cycle. Some legal scholars have called upon companies to follow an approach grounded in the Universal Declaration of Human Rights (UDHR) to provide a clear strategy against discriminatory impacts\cite{49}. While ethical frameworks and principles offer a roadmap, most are self-imposed, self-regulated, and challenging to implement, especially mathematically (e.g., fairness). However, in cases where AI-assisted decision-making technologies are deployed in social assistance systems, ethics are insufficient. Instead, there is a significant and dire need for actionable interventions that may include third-party oversight and accountability. 
\section{For further study}
The mobile health market is booming, and the potential of mHealth apps to improve access to real-time monitoring and health care resources for changing health-related behaviors is well established\cite{50}. Nevertheless, mHealth apps also pose problems concerning data privacy and potentially exacerbate unwanted bias, which can disproportionately impact specific populations. The more general question of "how can we mitigate biases in AI-based health care to ensure it improves care, rather than augment existing health disparities?" is at the center of this inquiry. Fortunately, there has been an increase in potential technical, normative, and policymaking strategies that help discover and address systematic biases that may manifest in direct-to-consumer health apps. It is also true, however, that no single bullet addresses the problem of unwanted bias in any intelligent system. From a lack of representative and complete data to implicit biases in various decision-making points within a technology's design, the problem of unwanted bias requires multiple stakeholders and iterative assessment. In addition, we highlight the following open questions for future inquiry:
\textbf{Business models and bias: }While we do not explicitly explore the intersection of business and unwanted bias, it should be noted that business models impact how systems are designed and implemented. In the case of the U.S. healthcare model, which is a multi-payer healthcare system, one obstacle to increasing digital health partnerships (between health professionals and digital health products) is the pricing and reimbursement structure\cite{51}. Therefore, it can be hypothesized that there is little financial incentive for either health professionals or digital health creators to work together. Thus, there is a need for a further understanding of the impact of the healthcare system's business model on the bias, a health apps' business model on the bias, and any implications around access and quality of access to digital health interventions.
\par
\textbf{Economic Feasibility of Bias Mitigation:} As more organizations enter the process of assessing, analyzing, and potentially mitigating bias in their intelligent systems, one outstanding unknown is the cost of engaging in the process. This includes both capacity and resource requirements that include designing auditing mechanisms as well as direct costs such as enhanced data collection (e.g., investing in new data gathering techniques or purchasing more datasets) as a result of any mitigation needs. 
\bibliography{sample}

\section{Author contributions statement}
K.G. and S.G. captured the policy and legal impact of AI in consumer mobile health technologies limitations,  N.R. analyzed and performed a literature survey on AI modeling bias and data collection bias on mhealth and applications alike within the consumer mobile health space. All authors reviewed the manuscript. 

\section{Additional information}

To include, in this order: \textbf{Accession codes} n/a; \textbf{Competing interests} none. 

The corresponding author is responsible for submitting a \href{http://www.nature.com/srep/policies/index.html#competing}{competing interests statement} on behalf of all paper authors. This statement must be included in the submitted article file.

Figures and tables can be referenced in LaTeX using the ref command, e.g. Figure \ref{fig:stream} and Table \ref{tab:example}.

\end{document}